\documentstyle[11pt]{article}
\newcommand{\E}{{\rm E}}
\newcommand{\Var}{{\rm Var}}
\newcommand{\Cov}{{\rm Cov}}
\newcommand{\Corr}{{\rm Corr}}

\newcommand \be  {\begin{equation}}
\newcommand \bea {\begin{eqnarray} \nonumber }
\newcommand \ee  {\end{equation}}
\newcommand \eea {\end{eqnarray}}
\begin{document}

\title{ \ \\ {\bf The Dynamics of the
Forward Interest Rate Curve with Stochastic String Shocks}\thanks{Thanks
to seminar participants at UCLA and Yale, and to Michael Blank, Michael
Brennan, Uriel Frisch, Mark Grinblatt,
Jon Ingersoll, Andrew Jeffrey, Olivier Ledoit, and Eduardo Schwartz, for
helpful conversations.}}

\author{\bf Pedro Santa-Clara\thanks{The Anderson Graduate School of
Management,
UCLA, 110 Westwood Plaza, Los Angeles, CA 90095-1481. Tel: (310) 206-6077.
E-mail: pedro.santa-clara@anderson.ucla.edu.}~ and Didier
Sornette\thanks{Institute of Geophysics and Planetary Physics, and
Department of
Earth and Space Sciences, UCLA, Los Angeles, CA 90095-1567; also at
CNRS and Universit\'e des Sciences, Parc Valrose, 06108 Nice Cedex 2, France.
Tel: (310) 825-2863.
E-mail: sornette@cyclop.ess.ucla.edu.}}

\date{University of California, Los Angeles \ \\ \ \\ {\normalsize First
Version:
March 1997 \ \\ This Version: \today}}

\maketitle

\begin{abstract}
\noindent
This paper offers a new class of models of the term structure of interest
rates. We allow each instantaneous forward rate to be driven by a different
stochastic shock, constrained in such a way as
to keep the forward rate curve continuous.
We term the process followed by the shocks to the forward 
curve ``stochastic
strings'', and construct them as the solution to stochastic partial
differential equations, that allow us to offer a variety of
interesting parametrizations.
The models can produce, with parsimony, any sort of correlation 
pattern among forward rates of different
maturities.
This feature makes the models consistent with any panel dataset of bond
prices, not requiring the addition of 
error terms in econometric models.
Interest rate options can easily be priced by simulation.
However, options can only be
perfectly hedged by trading in bonds of all maturities available.
\end{abstract}

\thispagestyle{empty}

\newpage

\pagenumbering{arabic}


\section{Introduction}


In this paper, we develop a new class of bond pricing models that greatly
extends the framework of Heath, Jarrow and Morton (HJM, 1992). Our
model is as parsimonious and
tractable as the traditional HJM model, but is capable of generating a much
richer class of dynamics and shapes of
the forward rate curve.
Our main innovation consists in having each instantaneous forward rate
driven by its own shock, while constraining these shocks in such a way as to
keep the curve continuous. This shock
to the curve is termed a ``stochastic
string'', following the physical analogy of a string whose shape changes
stochastically through time.

Using stochastic strings as noise source for the dynamics of the forward
curve has important economic advantages
over previous approaches. Existing term structure models have the same set of
shocks affect all forward rates. This
feature constrains the correlations between bond prices, and, therefore, the
set of admissible shapes and dynamics of
the yield curve. In term structure models with state variables (e.g. Vasicek,
1977 and Cox, Ingersoll and Ross, CIR,
1985), having complex shapes of the forward curve and corresponding elaborate
dynamics requires the introduction
of a large number of state variables. This makes the models very
unparsimonious and virtually impossible to
estimate. A consequence of the same problem is the difficulty that these
models have in fitting the term structure at
any given point in time, for given parameters estimated from a previous time
series of data.

Similar problems affect the traditional HJM model, where fitting closely
the initial term structure imposes
strong constraints on the dynamics of the forward rates, that are in general not
verified as the model is successively fitted
through time.

In contrast to these models, in our approach, any (finite) set of bond prices
is imperfectly correlated. The model is
thus fully compatible with any given panel dataset of bond prices, or, to put it differently, any
(finite) set of bond prices observed at some
(finite) sampling frequency is consistent with the model. There is thus no
need to add observation noise when estimating the model.

Another interesting characteristic of our approach is that, in general, it is
necessary to use a portfolio with an infinite
number of bonds to replicate interest rate contingent claims.
However pricing remains simple: interest rate options can in general 
be priced by simulation and, in some 
cases, in closed form. As with HJM, our approach does not allow the formulation of a 
partial differential equation to price derivatives.

We provide a detailed treatment of stochastic strings, and their stochastic
calculus. The framework we use is that of stochastic partial differential
equations (SPDE's) and our main tool is the calculus of Dirac distributions.
Methodologically, we attempt to present the results and their derivations
in the simplest and most intuitive way, rather than emphasize mathematical
rigor.

Kennedy (1997) and, in recent independent work, Goldstein (1997), propose a
similar approch to modeling forward rates. Kennedy (1997) simply models the
forward rate curve as a Gaussian random field. Goldstein (1997) uses a model
similar to ours, letting the forward
rate curve be shocked by two of the strings that we analyse as special
examples of our framework.

The paper is organized as follows. In the next section we define the
primitives of our approach and solve a
simple model where forward rates are driven by Brownian motion, to compare
with the models driven by stochastic
strings. Section 3 defines string shocks and their properties and
offers the general no-arbitrage condition for forward
rate dynamics driven by stochastic strings. Section 4
discusses te construction of
stochastic strings as solutions of SPDE's. In section 5,
we present a collection of examples of
stochastic strings, that present interesting properties. 
Section 6 presents a discussion of the emprical implications of the model,
in comparison with traditional factor models.
In section 7, we show how interest rate derivatives can
be priced and hedged in the model. Section 8 concludes with some directions for future work.


\section{The Traditional Model: Brownian Motion as Noise Source}


\label{simpmod}

We postulate the existence of a stochastic discount factor (SDF) that
prices all assets in this economy, and denote it by $M$.\footnote{This
process is also termed the pricing kernel, the
pricing operator, or the state price density. We use these terms
interchangeably. See Duffie (1996) for the theory
behind SDF's.} This SDF can be thought of as the nominal, intertemporal,
marginal rate of substitution for
consumption of a representative agent in an exchange economy. It is well
known that assuming that no dynamic
arbitrage trading strategies can be implemented by trading in the financial
securities issued in the economy is
roughly equivalent to the existence of a strictly positive SDF. For no
arbitrage opportunities to exist, and under an
adequate definition of the space of admissible trading strategies, the
product $MV$ must be a martingale, where
$V$ is the value process of any admissible self-financing trading strategy
implemented by trading on financial
securities. Then,
\be V(t)=\E_t \left[V(s) \frac{M(s)}{M(t)} \right]~, \label{martcond} \ee
where $s$ is a future date and $\E_t
\left[x\right]$ denotes the mathematical expectation of $x$ taken at time
$t$.
In particular, we require that a bank account and zero-coupon discount bonds
of all maturities satisfy this condition.

A security is referred to as a (floating-rate) bank account, if it is
``locally riskless''.\footnote{A security is ``locally
riskless'' if, over an instantaneous  time interval, its value varies
deterministically. It may
still be random, but there is no Brownian  term in its dynamics.} Thus, the
value at time $t$, of an initial investment
of $B(0)$ units in the bank account that is continuously reinvested, is given
by
\be B(t) = B(0)\exp
\left\{ \int_{0}^{t}r(s)ds \right\}~, \ee where $r(t)$ is the instantaneous
nominal interest rate.

We further assume that at any time $t$ riskless discount bonds of all {\it
maturity dates} $s$ trade in this economy and let $P(t,s)$ denote the time
$t$ price of the $s$ maturity bond. We
require that $P(s,s)=1$, that $P(t,s)>0$ and that $\partial P(t,s)/\partial
s$ exists.

Instantaneous forward rates at time $t$ for all {\it times-to-maturity}
$x>0$, $f(t,x)$, are defined by 
\be f(t,x) = -
\frac{\partial \log P(t,t+x)}{\partial x}~,  
\ee 
which is the rate that can be
contracted at time $t$ for instantaneous
borrowing or lending at time $t+x$. We require that the initial forward curve
$f(0,x)$, for all $x$, be continuous.

Equivalently, from the knowledge of the instantaneous forward rates for all
times-to-maturity between $0$ and time $s-t$, the price at time $t$ of a bond
with maturity $s$ can be obtained by
\be P(t,s) = \exp \left\{ - \int_0^{s-t}
dx f(t,x)   \right\} ~. \label{intfor} \ee
Forward rates thus fully represent the information in the prices of all
zero-coupon bonds.

The spot interest rate at time $t$, $r(t)$, is the instantaneous forward
rate at time $t$ with time-to-maturity $0$,
\be r(t) = f(t,0) ~. \label{rate} \ee

We use forward rates with fixed {\it time-to-maturity} rather than fixed {\it
maturity date}.\footnote{The model of
HJM starts from processes for forward rates with a fixed maturity date. This
is different from what we do. If we use
a ``hat'' to denote the forward rates modeled by HJM,
\[ \hat{f}(t,s)=f(t,s-t) \]
or, equivalently, \[ f(t,x)=\hat{f}(t,t+x) \] for fixed $s$. Musiela (1993)
and Brace and Musiela (1994) define
forward rates in the same fashion. Miltersen, Sandmann and Sondermann (1997),
and Brace, Gatarek and Musiela
(1995) use definitions of forward rates similar to ours, albeit for
non-instantaneous forward rates.}
Modelling forward rates with fixed time-to-maturity is more
natural for thinking of the dynamics of the entire forward curve as the
shape of a string evolving in time. In contrast, in HJM, forward rate
processes disappear as time reaches their
maturities. Note, however, that we still impose the martingale condition on
bonds with fixed {\it maturity date},
since these are the financial instruments that are actually traded.

We now introduce a simple model of forward rate dynamics, which is a
variation of HJM introduced by Musiela
(1993) and Brace and Musiela (1994), both for completeness and as a benchmark
against which to check the models
in later sections.

Let us first simply model all forward rates by
\be df(t,x) = \alpha(t,x) dt +
\sigma(t,x)dW(t) ~. \label{fordyn} \ee
or, in integral form,
\be f(t,x) =
f(0,x)
+ \int_0^t dv~ \alpha(v,x) + \int_0^t dW(v)~\sigma(v,x) \label{fordynint} \ee

From the definition of bond prices, we get at fixed $s$ that
\be d \log P(t,s) =
f(t, s-t)~dt - \int_0^{s-t} dy ~df(t,y) ~. \ee
With Ito's Lemma, we can calculate the dynamics of the prices of bonds with
fixed maturity date $s$,
\be \frac{dP(t,s)}{P(t,s)} = \left[ f(t,s-t) -
\int_0^{s-t}dy~ \alpha(t,y) + \frac{1}{2}\left(\int_0^{s-t} dy~\sigma(t,y)
\right)^2  \right]dt - \left(\int_0^{s-t} dy ~\sigma(t,y)  \right)dW(t)
\label{bpri} \ee

Finally, we postulate the following dynamics for the SDF\footnote{$M$ can
have other forms than (\ref{sdf}). The
extension to several Brownian motion terms is trivial, along the lines of
HJM. More importantly, the SDF can have
jumps  - that is a Poisson process term - due to discrete changes in the
arrival of information.}
\be \frac{dM(t)}{M(t)}  = -r(t)~ dt  + \phi(t)~dW(t) ~. \label{sdf} \ee
The drift of $M$ is justified from the well-known martingale condition on the
product of the bank account with the
SDF. The process
$\phi$ denotes the market price of risk, as measured by the covariance of
asset returns with the SDF.

The no-arbitrage condition for buying and holding bonds implies that $PM$ be
a martingale in time, for any bond
price $P$. Technically, this
amounts to imposing that the drift of $PM$ be zero,
\be - r(t) + f(t,s-t) - \int_0^{s-t}dy ~\alpha(t,y)
+ \frac{1}{2}\left(\int_0^{s-t}dy ~\sigma(t,y)  \right)^2 -
\phi(t)\left(\int_0^{s-t}dy ~\sigma(t,y)  \right)=0~. \ee
Equivalently, for all $t$ and $s$, with $x\equiv s-t$
\be f(t,x) = r(t) + \int_0^{x} dy~ \alpha(t,y) -
\frac{1}{2}\left(\int_0^{x}dy ~\sigma(t,y)  \right)^2 +
\phi(t)\left(\int_0^{x}dy ~\sigma(t,y)  \right)
\label{erfdvxb}
\ee

We can differentiate this no-arbitrage condition with respect to
$x$, and obtain
\be
\alpha(t,x) = \frac{\partial }{\partial x}\left\{ f(t,x) +
\frac{1}{2}\left(\int_0^{x}dy~ \sigma(t,y)  \right)^2 -
\phi(t)\left(\int_0^{x}dy~
\sigma(t,y)  \right) \right\} ~,
\ee
or,
\be \alpha(t,x) = {\partial f(t,x)
\over \partial x} +  \sigma(t,x) \biggl( \int_0^{x} dy~\sigma(t,y)  - \phi(t)
\biggl) ~.
\ee
where ${\partial f(t,x)  \over \partial x}$ is the slope of the
forward curve, at time $t$ for time-to-maturity $x$. This no-arbitrage
condition clearly shows that the diffusion
function $\sigma$ must go to zero when the maturity date goes to
infinity to ensure the finiteness of $\alpha$.\footnote{See Dybvig,
Ingersoll and Ross
(1996) and Jeffrey (1997).}
The term ${\partial f(t,x) \over \partial x}$ stems from the parametrization
in terms of the {\it time-to-maturity}. It
would be absent in the usual HJM fixed time-of-maturity parametrization. Let us
stress that the two formulations are completely equivalent.

We thus get the following arbitrage-free model of forward rate dynamics
\be
d_t f(t,x) = \biggl({\partial f(t,x)  \over \partial x} + a(t,x) \biggl) dt +
\sigma(t,x)~dW(t) ~, \label{fordynsol} \ee
denoting
\be a(t,x) = \sigma(t,x) \biggl( \int_0^{x} dy~\sigma(t,y)  - \phi(t) \biggl)
\label{sghroittu}
\ee
This condition can also be found in Musiela (1993) and Brace and Musiela
(1994).

We see that, at some date $t$, estimating the drift and volatility functions
of any forward rate, along with the slope
of the forward rate curve at that point, recovers the market price of risk
process $\phi$.\footnote{This extends
naturally to the case where there is a finite number of Brownian increments
in the SDF and the forward rate
processes. In this case, we would need information about the drifts and
volatilities of as many forward rates as
Brownian increments in the model.} The use of a SDF for pricing is just a
useful presentation device. The no-
arbitrage condition for a single forward rate allows us to construct the
dynamics of all forward rates from the
knowledge of their volatilities.

In appendix A, we solve equation (\ref{fordynsol}) by first rewritting it as
\be {\partial f(t,x)  \over \partial t} - {\partial f(t,x)  \over \partial x}
= a(t,x) + \sigma(t,x)~{dW(t) \over dt}~,
\label{zetr}
\ee
where ${dW(t) \over dt}$ is such that, when integrated over time from $0$ to
$t$, we obtain $\int_0^t dv~{dW(v) \over dv} = \int_0^t dW(v)$.
\footnote{In the equation, ${dW(t) \over dt}$ is
an alternative notation for the noise term $\eta$ defined later in (\ref{browhdfd}) with
covariance (\ref{cvdgdggh}).} We show
that the solution of (\ref{zetr}) is
\be
f(t,x) = f(0,t+x) + \int_0^t dv~ a(v,t+x-v) + \int_0^t dW(v) ~
\sigma(v,t+x-v)~.
\label{lkjbc}
\ee
Notice that if $\sigma(t,x) = 0$ (no fluctuations), we recover
the natural no-arbitrage condition $f(t,x) = f(0,t+x)$ stating that the
instantaneous forward rate is the same at time $0$ and at time $t$ for the
same maturity date $t+x$.


\section{A Better Model: Stochastic String Shocks}


\label{betmodel}

In this section we develop a more general model of the forward rate curve
driven by a stochastic string.
Instead of the traditional model (\ref{fordyn}), we now
model the dynamics of the forward rates by
\be
d_t f(t,x) = \alpha(t,x) dt + \sigma(t,x) d_t Z(t,x)~,
\label{modelsheet}
\ee
or, in integral form, by
\be
f(t,x) = f(0,x) + \int_0^t dv~\alpha(v,x)  + \int_0^t d_v Z(v,x)~\sigma(v,x)
~.
\ee
where the stochastic process $Z(t,x)$ generalizes to two dimensions the
previous one-dimensional Brownian motion $W(t)$. The important innovation in
(\ref{modelsheet}) is that the
stochastic process depends not only on time $t$ but also on time-to-maturity
$x$. The notation
$d_t Z(t,x)$  denotes a stochastic perturbation to the forward rate curve
at time $t$, with different magnitudes for forward rates with different
times-to-maturity. It is straightforward to extend the model to include more
than one string shock or to combine string shocks with Brownian motion 
shocks.

We stress that (\ref{modelsheet}) is not the
infinite-dimensional generalization of the multi-factor HJM model in which
all forward rates are subjected to the same (possibly infinite-dimensional)
set of stochastic processes. We have one stochastic
process per time-to-maturity. Our process
is infinite dimensional in the sense that the set of times-to-maturity 
in the continuous description of the yield curve
has the order of infinity of the continuum.

We impose several requirements on $Z$ to qualify as a string shock to the
forward rate curve\,:
\begin{enumerate}
\item $Z(t,x)$ is continuous in $x$ at all times $t$;
\item $Z(t,x)$ is continuous in $t$ for all $x$;
\item The string is a martingale in time $t$, $\E \left[ d_t Z(t,x) \right]=0$, for all $x$;
\item The variance of the
increments, $\Var \left[ d_t Z(t,x) \right]$, does not depend on $t$ or $x$;
\item The correlation of the increments,  $\Corr \left[ d_t Z(t,x),  d_t Z(t,x') 
 \right]$, does not depend on $t$. 
\end{enumerate}

We will see that the first two conditions are automatically satisfied by
taking $Z$ to be the solution of a SPDE with
at least one partial derivative in $x$ and $t$. 
Condition 3 captures the unforecastable aspect of shocks. Condition 4 corresponds to
making all shocks affecting the forward curve have the same intensity.
This intensity can then be changed, or
``modulated'', by the function $\sigma(t,x)$ taking different values at
different times to maturity $x$. The final condition imposes that the correlation between 
shocks two forward rates of different maturities 
depend only on these maturities.
Conditions 3, 4 and 5 together make the strings Markovian.
We will see
that all string shocks produced as solutions of SPDE's
have Gaussian distributions, which are completely characterized by the first two moments.

To have each forward rate, for each time-to-maturity, be driven by its
own noise process, we could simply use infinite-dimensional Brownian motion,
using a
simple Brownian motion as a shock to each forward
rate, independent from the Brownian motion used to perturb
forward rates with different times-to-maturity. However, such a model makes
the forward rate curve discontinuous
as a function of time-to-maturity,
so that, over time, any two forward rates can become very distant from each
other. Although a discontinuous
forward curve violates no arbitrage conditions, it is intuitively 
unlikely.

We develop the model, assuming the existence of a string shock that satisfies our
requirements. In the next sections, we
show how such string shocks can be constructed and examine several parametric
examples.

We now take the process for the SDF given by
\be \frac{dM(t)}{M(t)}  =-r(t)~
dt  + \int_0^{\infty}~ dy ~\phi(t,y)~ d_t Z(t,y) \label{sdf2} \ee
where the pricing kernel is driven by a ``weighted sum'' of the shocks that
affect the forward rate curve, and the market prices of risk can in principle
be different for each shock to the curve.

We follow the steps of the derivation of the no-arbitrage
condition on the drift of forward rates of section \ref{simpmod}.
We have
\be
y(t,s)\equiv d_t \log P(t,s) = \left[ f(t,s-t) - \int_0^{s-t} \alpha(t,y)~dy
\right]dt
-  \int_0^{s-t} dy ~\sigma(t,y) ~d_t Z(t,y)~.
\label{bpdfri}
\ee

We need the expression of $d P(t,s) / P(t,s)$ which is obtained from
(\ref{bpdfri}) using Ito's calculus. In order to get Ito's term in the drift,
recall that it results from the fact that, with
$y$ stochastic,
\be
d_t F(y) = {\partial F \over
dy} d_t y + {1 \over 2}  {\partial^2 F \over dy^2} \Var \left[ d_t y
\right]
\ee
so that the differential $d_t F$ up to order $dt$ contains contributions from
all the covariances of the form
$\Cov \left[ d_t Z(t,x),d_t Z(t,x')
\right]$ appearing in $\Var \left[ d_t y(t,s) \right]$.

Taking the expectation of the square of the stochastic term in the r.h.s. of
(\ref{bpdfri}), bond price dynamics can be
written as
$$
{d P(t,s) \over P(t,s)} = \left[ f(t,s-t) - \int_0^{s-t} \alpha(t,y)~dy + {1
\over 2} \int_0^{s-t}dy\int_0^{s-t}dy' c(t,y,y')\sigma(t,y) \sigma(t,y')
\right]dt
$$
\be
-  \int_0^{s-t} dy ~\sigma(t,y) ~d_t Z(t,y)~.
\label{refnnn}
\ee
where
\be
c(t,y,y') ~ \equiv \Corr \left[ d_t Z(t,y),d_t Z(t,y') \right]~.
\label{coroduu}
\ee
and we make use of $\Var \left[ d_t Z(t,x) \right] = dt$.
In the next sections, we offer several examples of stochastic strings
and the correlation functions of their time
increments.

The no-arbitrage condition becomes
$$
- r(t) + f(t,x) - \int_0^{x} dy~ \alpha(t,y) + {1 \over
2} \int_0^{x}dy\int_0^{x}dy'~ c(t,y,y') \sigma(t,y) \sigma(t,y')
$$
\be
-  \int_0^{\infty}dy\int_0^{x}dy' ~c(t,y,y') \phi(t,y)
\sigma(t,y')
= 0
\label{maringhdk}
\ee
that we can differentiate with respect to $x$,\footnote{In principle,
differentiating (\ref{maringhdk}) is only
warranted if $f(t,x)$ is differentiable in $x$. This is the case for some
strings constructed below, such as the
integrated O-U sheet, but is not true for others, such as the O-U sheet. In
these cases, ${\partial f(t,x) \over \partial
x}$ is not a function but is similar to an increment of a Brownian motion.
However, the results below still hold,
by reading ${\partial f(t,x) \over \partial x}$ as a  convenient notation 
for the formal integral calculus used.} to
obtain
\be
\alpha(t,x) =
\frac{\partial f(t,x)}{\partial x} +
\sigma(t,x) \left( \int_0^{x}dy~ c(t,x,y) \sigma(t,y)
- \int_0^{\infty}dy~ c(t,x,y) \phi(t,y) \right)
\label{eqkdkkd}
\ee

If $c(t,x,y)$ in (\ref{eqkdkkd}) is one, we recover the traditional model of
a single Brownian motion.

We thus get the following arbitrage-free model of forward rate dynamics
$$
d_t f(t,x) = \left[\frac{\partial f(t,x)}{\partial x} +
\sigma(t,x) \left( \int_0^{x}dy~ c(t,x,y) \sigma(t,y)
- \int_0^{\infty}dy~ c(t,x,y) \phi(t,y) \right) \right]dt
$$ \be
+ \sigma(t,x) d_t Z(t,x)~.
\label{bphhbcgg}
\ee

Expression (\ref{bphhbcgg}) can be solved in a similar way to section
\ref{simpmod}. We denote
\be
A(t,x) = \sigma(t,x) \left( \int_0^{x}dy~ c(t,x,y) \sigma(t,y)
- \int_0^{\infty}dy~ c(t,x,y) \phi(t,y) \right) ~.
\label{aaaaqqzzzs}
\ee
to finally obtain
\be
f(t,x) = f(0,t+x)  + \int_0^t dv~ A(v,t+x-v)
+ \int_0^t d_v Z (v,t+x-v) ~ \sigma(v,t+x-v)  ~.
\label{lkjdgbc}
\ee

We see that the correlation function of the time increments of the
strings is sufficient (together with the risk premia) to characterize
arbitrage-free dynamics of the forward curve.

We can still invert the forward curve to obtain the market prices of risk
needed for pricing derivatives. However, we now need to use the entire 
forward rate curve rather than a smal number of forward rates (equal to 
the number of Brownian motions driving the curve in the traditional model). 
Note that if we assume that $\phi(t,x) = \phi(t)$,
thus constraining the SDF (\ref{sdf2}), the ``inversion'' of the 
forward rate curve to extract the market prices of
risk becomes much easier.


\section{Construction of Stochastic String Shocks}


This section shows how to construct stochastic string shocks that
can be used to perturb the forward rate curve.

We start by reviewing the construction of Brownian motion from a probabilistic
point of view and then as the solution to a {\it stochastic ordinary differential
equation} (SODE). This allows us to obtain a natural generalization of Brownian
motion to string
motion as the solution of {\it stochastic partial differential
equations} (SPDE's). We then formulate the solutions to these SPDE's 
in terms of the Green function of the equation and find constraints on 
this function to impose our requirements for string shocks.

\subsection{Brownian motion as solution of a SODE}

The usual description of stochastic processes in Finance is based on the
Brownian motion $W$ defined by
\be W(t) = \int_0^t d_u W(u) ~, \ee
normally distributed with
\be \E
\left[d_t W(t) \right] = 0~,
\ee
and
\be \Var \left[ d_t W(t) \right] = dt~.
\label{ezddscxfgws} \ee

Alternatively, $W$ can be defined as the solution of
the following SODE\,:
\be {d W(t) \over dt} = \eta (t)~.
\label{browhdfd}
\ee
where $\eta$ is white noise, characterized by the covariance
\be
\Cov \left[ \eta (t),\eta (t')
\right] = \delta (t-t')~,
\label{cvdgdggh}
\ee
and $\delta$ designates the Dirac function.\footnote{The Dirac function
is a distribution in the sense of
Schwartz's theory of distributions, or
generalized functions. It is such that, for arbitrary $f(x)$,
$\int_{x_1}^{x_2} f(x) \delta(x-x_0)
= f(x_0)$ if $x_1 < x_0 < x_2$, ${1 \over 2} f(x_0)$ if
$x_1 = x_0$ or $x_0 = x_2$ and zero otherwise.
$\delta(x)$ is obtained as the limit of functions when their width goes to zero
and their height goes to infinity in such a way that their integral remains one. An
example
is the Gaussian $\lim_{\sigma \to 0}
{1 \over \sqrt{2\pi}~\sigma} e^{-{x^2 \over 2 \sigma^2}} = \delta(x)$.}

The SODE (\ref{browhdfd}) describes a particle at position $W$ which is
incessantly subjected to random forces
leading to random variations $\eta$ of its velocity.

The solution of (\ref{browhdfd}) is formally
\be W(t) =
\int_0^t dv ~\eta (v)~. \label{erfvdfdfq} \ee
which shows that $dt ~\eta (t)$ is
simply a notation for $d_t W(t)$, and, as usual, has rigorous mathematical
meaning only under the integral representation. From (\ref{erfvdfdfq}), we
see (by the Central Limit Theorem) that
$W$ will be Gaussian, with mean zero, and we can calculate easily the
covariance
\be
\Cov \left[ W (t) , W (t') \right] = \int_0^t
dv \int_0^{t'} dv' ~\Cov \left[ \eta (v) ,~\eta (v') \right] = \int_0^t dv
\int_0^{t'}
dv'~\delta(v-v') = (t \wedge t') ~,
\label{Browojhhfhhddkl}
\ee
where $(t \wedge t')$ stand for $\min(t,t')$.

Finally, we define $d_t W(t)$ as the limit of $[W(t+ \delta t) - W(t)]$ when
the small but finite increment of time
$\delta t$, typical of what we have to deal with in real time series, becomes
the infinitesimal $dt$. Using
(\ref{Browojhhfhhddkl}), we get $\Var \left[ W(t+ \delta t) - W(t) \right] =
\delta t$ which recovers (\ref{ezddscxfgws}) in the infinitesimal time
increment limit.

The definition of the Brownian motion as the solution of a SODE is very
useful to generalize to other processes. Maybe the simplest extension is the
well-known Ornstein-Uhlenbeck (O-U)
process $U(t)$ which can be defined as the solution of the following SODE\,:
\be {d U(t) \over dt} = -\kappa U(t) + \eta (t)~, \label{ejjfppoi} \ee
where $\kappa$ is a positive constant. In addition to the
random variations $\eta$ of the velocity, the Brownian particle is now
subjected to a restoring force tending to bring it back to the origin, i.e.\
mean reversion.
The formal solution reads
\be
U(t) = \int_0^t dv~\eta(v) ~e^{-\kappa(t-v)}~.
\ee
Its covariance is
\be
\Cov \left[  U(t),~ U(t') \right] = {1 \over 2\kappa} \biggl(
e^{-\kappa |t-t'|} - e^{-\kappa (t+t')}\biggl) ~,
\ee
which goes to ${1 \over 2\kappa} e^{-\kappa |t-t'|}$ at large
times.\footnote{Using $2 ~(t \wedge t') = t+t' - |t-t'|$.}
By adding more terms in (\ref{ejjfppoi}), more complex stochastic processes
can be easily generated.

\subsection{Stochastic strings as solutions of SPDE's}

We now generalize the stochastic processes driving the uncertainty in the
forward rate curve from solutions of
SODE's to solutions of SPDE's.
In the context of differential calculus, this is the most natural and general
extension that can be
performed.\footnote{Further extensions will include fractional differential
equations and integro-differential
equations, including jump processes.}
The introduction of partial
differential equations is called for in order to account for the continuity
condition and to deal with the two variables, time and time-to-maturity, and
their interplay.

Brownian motion depicts, in physical terms, the motion of a particle
subjected to random velocity variations. The
analogous physical system described by SPDE's is a {\it stochastic string}.
In our application of stochastic strings as
the noise
source in the dynamics of the forward rate curve, the length along the
string is the time-to-maturity, and the string configuration gives the
amplitude of the shocks at a given time, for each
time-to-maturity. The
set of admissible dynamics of the configuration of the  string as a function
of time depends on the specification of
the SPDE.

In the present paper, we restrict our attention to linear SPDE's,
in which the highest derivative is, in most cases, second order. This second
order derivative has a simple physical
interpretation\,: the string is
subjected to a tension, like a piano chord, that tends to bring it back to
zero deformation. This tension forces the
``coupling'' among different times-to-maturity so that the forward rate curve
remains continuous. In principle, a more general formulation would 
consider SPDE's with terms of arbitrary
derivative orders.\footnote{Higher order
derivatives also have an intuitive physical interpretation. For instance,
going up to fourth order derivatives in the
SPDE correspond to the dynamics of a {\it beam}, which has bending elastic
modulus tending to restore the beam
back to zero deformation, even in absence of tension.} However, it is easy to
show that the tension term is the
dominating restoring force, when present, for deformations of the string
(forward rate curve) at long ``wavelengths''
along the time-to-maturity axis. Second order SPDE's are thus generic in the
sense of a systematic
expansion.\footnote{There are situations where the tension can
be made to vanish (for instance in the presence of a rotational symmetry) and
then the leading term in the SPDE
becomes the fourth order ``beam'' term.}

The general form of second-order linear SPDE's reads
\be a(t,x) {\partial^2 X(t,x)
\over \partial t^2} + 2 b(t,x) {\partial^2 X(t,x) \over \partial t \partial
x} +
c(t,x) {\partial^2 X(t,x) \over \partial x^2} = f\left(t,x, X(t,x), {\partial
X(t,x) \over
\partial t}, {\partial X(t,x) \over \partial x}\right)~,
\label{zz1} \ee
where $X$ is the stochastic string shock we want to
characterize. In the present paper, we restrict
our discussion to {\it linear} equations, where $F$ has the form
\be f\left(t,x, X(t,x), {\partial X(t,x) \over
\partial t}, {\partial X(t,x) \over \partial x}\right) = d(t,x) {\partial
X(t,x) \over
\partial t} + e(t,x) {\partial X(t,x) \over \partial x} + g(t,x) X(t,x) +
s(t,x)~.\label{zz2} \ee
$s$ is the ``source'' term that will be generally taken to
be Gaussian white noise $\eta $ characterized by the covariance
\be
\Cov \left[ \eta (t,x),~ \eta (t',x') \right] = \delta (t-t') ~\delta
(x-x')~,
\label{cvgherfv}
\ee
where $\delta$ denotes, as before, the Dirac distribution.
Expression (\ref{zz1}) together with (\ref{zz2}) is the most general
linear second-order SPDE in two variables. The solution $X$ exists
and its uniqueness is warranted once ``boundary'' conditions are given, such
as, for instance, the initial value of the
string $X(0,x)$, for all $x$, as well as any constraints on the particular form of equation
(\ref{zz1}).\footnote{See for instance
Morse and Feshbach (1953).}

\subsection{Green function formulation}

All solutions of the above linear SPDE can be characterized by
\be
X(t,x) = X(0,x) + \int_0^t dv \int_{-\infty}^{\infty} dy
~G(t,x|v,y)~ \eta(v,y)~,
\label{dggvvgggg}
\ee
where the Green function $G$ contains all the information in the
underlying SPDE and thus of the coupling between the different 
times-to-maturity $x$.

Consider a single impulsive source term $\eta(v,y) = \delta(v-t_0)
\delta(y-x_0)$. Then,
\be
X(t,x) = X(0,x) + G(t,x|t_0,x_0). \ee
The
Green function $G$ thus describes the deterministic evolution in the future
of the process $X$ at all times-to-maturity due to an impulsive
perturbation that occurs at time $t_0$ at the time-to-maturity $x_0$.
It thus embodies all the coupling between different
times-to-maturities.

We have
\be
\Var \left[ X(t,x)\right] =  \int_0^t dv \int_{-\infty}^{\infty} dy
~[G(t,x|v,y)]^2~,
\label{daapppss}
\ee
and
$$
\Cov \left[ d_t X(t,x),d_t X(t,x') \right] = \lim_{\delta t
\to dt}
\int_t^{t+\delta t} dv \int_{-\infty}^{\infty} dy ~G(t+\delta t,x|v,y)
~G(t+\delta t,x'|v,y)
$$
\be
+ \int_0^t dv \int_{-\infty}^{\infty} dy [G(t+\delta t,x|v,y) -G(t,x|v,y)]
[G(t+\delta t,x'|v,y) -G(t,x'|v,y)]
\label{daaaasss}
\ee

However, this general solution does not satisfy all our conditions for a
stochastic string shock to the forward curve. We want to impose in addition
that
$X(t,x)$ is a martingale, and that
the variance of its increments, $\Var \left[ d_t X(t,x) \right]$, does not depend on $x$.
In appendix B, we show that strings that satisfy our requirements
can be written as
\be
Z(t,x) = Z(0,x) + \int_0^t dv ~\int_{x_0}^{j(x)} dy
~h(v,x,y) \eta(v,y)~,
\label{dzsdsd}
\ee
where $g$, $j$ and $h$ are functions to be constrained in order to have the variance 
of the increments independent of $x$, and
the lower bound can be adjusted in order to get rid of convergence
problems at infinity. This leads to
\be
d_t Z(t,x) = dt~
\int_{x_0}^{j(x)} dy ~h(t,x,y) \eta(t,y)~,
\label{dzzzzssd}
\ee
so that
\be
\Var \left[ d_t Z(t,x) \right] = dt~ \int_{x_0}^{j(x)} dy
~[h(t,x,y)]^2 ~.
\label{dzzzfqdfqdfqsdf}
\ee
The additional condition that $\Var \left[ d_t Z(t,x) \right]$ does not depend on $x$ leads to
several solutions.

A first class of solutions is obtained if we assume that $h(t,x,y)$ is
independent of $y$.
The integration gives then simply $j(x) [h(t,x)]^2$. Equating to a constant
(equal to $1$ without loss of generality), this gives
\be
h(x) = {1 \over \sqrt{j(x)}}~.
\ee
The corresponding stochastic string is thus defined by
\be
d_t Z(t,x) = dt~ {1 \over \sqrt{j(x)}} \int_0^{j(x)} dy~\eta(t,y)~.
\label{gindep}
\ee
and the correlation of the increments is
\be
c(t,x,x') \equiv \Corr \left[ d_t Z(t,x), d_t Z(t,x') \right] =
\sqrt{j(x) \over j(x')} ~,
\label{rfdscxw}
\ee
if $j(x) < j(x')$. The role of $x$ and $x'$ in (\ref{rfdscxw}) are inversed
if $j(x) > j(x')$. The corresponding SPDE reads
\be
{\partial \over \partial x} \biggl(\sqrt{j(x)} {\partial Z(t,x)
\over \partial t}\biggl)
= \sqrt{d j(x)\over dx} ~\eta(t,x)~.
\ee

An alternative class of solutions can be obtained by noting that
\be
[h(x)]^2 = {1 \over x} {d \over dy}[j]^{-1}(y)~,
\label{dhhhhwxxwxwxx}
\ee
where $[j]^{-1}$ is the inverse function of $j$ defined by
$j([j]^{-1}(y))=y$,
verifies the constraint that $\Var \left[ d_t Z(t,x) \right]$ be
independent of $x$.
This corresponds to
\be
d_t Z(t,x) = dt~ {1 \over \sqrt{x}} \int_0^{j(x)} dy~
\biggl({d \over dy}[j]^{-1}(y)\biggl)^{1/2} ~\eta(t,y)~.
\ee

A generalization of (\ref{dhhhhwxxwxwxx}) is
\be
[h(x)]^2 = {1 \over x^{\alpha}} {d \over
dy}\biggl([j]^{-1}(y)\biggl)^{\alpha}~,
\label{dhhhhwaaaawxx}
\ee
where $\alpha >0$ is arbitrary. In fact, even more generally, consider an
arbitrary
monotonous function $l$. Then,
\be
[h(x)]^2 = {1 \over l(x)} {d \over dy}l\biggl([j]^{-1}\biggl)(y)~
\label{dhhhzznnaawxx}
\ee
satisfies the constraint that $\Var \left[ d_t Z(t,x) \right]$
does not depend on $x$. This leads to
\be
d_t Z(t,x) = dt~ {1 \over \sqrt{l(x)}} \int_0^{j(x)} dy ~
\sqrt{{d \over dy}l\biggl([j]^{-1}\biggl)(y)}~ \eta(t,y)~.
\ee
The correlation of the increments is
\be
c(t,x,x') \equiv \Corr \left[ d_t Z(t,x), d_t Z(t,x') \right] =
\sqrt{l(x) \over l(x')} ~,
\label{rfdscxwwwww}
\ee
if $j(x) < j(x')$. The role of $x$ and $x'$ in (\ref{rfdscxwwwww}) are
inversed
if $j(x) > j(x')$. This provides a slight
generalization to (\ref{rfdscxw}) since a different function appears in the
correlation function and in the inequality condition on $x$ and $x'$.

The corresponding SPDE is
\be
{\partial \over \partial x} \biggl(\sqrt{l(x)} {\partial Z(t,x) \over
\partial t}
\biggl)
= \sqrt{d j(x)\over dx}~\sqrt{{d l(x)\over dx}}~\eta(t,x)~.
\ee

\subsection{Differentiability in time-to-maturity}

All the processes studied above are continuous both in $t$ and $x$ but
not differentiable in either $t$ or $x$. Of special interest is the
non-differentiability in $x$.\footnote{We show in appendix C how to determine
the differentiability of string shocks.}

We would like now to discuss a class of processes which are
non-differentiable in time but differentiable in time-to-maturity, while still keeping
an infinite set of shock processes, one for each time-to-maturity
$x$.\footnote{HJM provides an obvious example of a stochastic process
which is non-differentiable in time and differentiable in $x$, but this is
a degenerate case as the same stochastic process drives all
times-to-maturities.} Economics has little to say with respect 
to the differentiability of forward rate curves. Furthermore, this 
issue cannot be resolved empirically, since forward rates of very 
close maturities cannot be observed. The use of string shocks that 
produce differentiable curves is thus fundamentally a matter of taste.

An intuitive strategy to obtain a shock differentiable in time-to-maturity
is to integrate any of the previously defined strings in the
$x$-variable\footnote{Smoother strings can be obtained by higher-order integration.}
\be
Y(t,x) = \int_{0}^x dy ~ Z(t,y)~.
\label{ervcpo2}
\ee
By definition, ${\partial Y(t,x) \over \partial x} = Z(t,x)$.
Since $Z(t,x)$ is continuous in $x$, ${\partial Y(t,x) \over
\partial x}$
is also continuous in $x$ and $Y(t,x)$ is thus differentiable with respect to
$x$.

Since $Z(t,y)$ satisfies a SPDE, we see that, by replacing
$Z(t,y)$ by ${\partial Y(t,x) \over \partial x}$ in this equation, that
$Y(t,x)$ is also the solution of a SPDE.

The correlation of the increments is
\be
c(t,x,x') \equiv \Corr \left[ d_t Y(t,x), d_t Y(t,x') \right] =
x~\bigg[\int_0^x \sqrt{l(y) \over l(x)} + \int_x^{x'} \sqrt{l(x) \over
l(y)} \biggl]~,
\label{rqsdfqdgfqdw}
\ee
for $x < x'$, assuming that $j(x)$ is monotonous increasing. The
corresponding expressions for other cases
are straightforwardly derived.

The variance is
\be \Var \left[ d_t Y(t,x) \right] =  dt~x ~\int_{0}^{x} dy~\sqrt{l(y) \over
l(x)}~,\ee
which is independent of $x$ if $l(x) = c_1 e^{c_2(x-1/x)}$, where $c_1$ and
$c_2$ are two arbitrary constants.

\section{Parametric Examples}

In this section we examine several examples of stochastic strings that
are solutions to SPDE's, and consider their interest as a noise
source in a model of the dynamics of the forward rate curve.

\subsection{The Brownian sheet}

We start by discussing the simplest example of a SPDE, which however does not
satisfy our criteria. The example is however useful as it 
is the basis for the derivation of other string processes.

The simplest SPDE in the class given by (\ref{zz1}) and
(\ref{zz2}) is
\be
{\partial^2 W(t,x) \over \partial t \partial x} = \eta (t,x)~,
\label{browhzd}
\ee
where $\eta (t,x)$ is a white noise both in time and $x$, characterized by
the covariance function (\ref{cvgherfv}).
By inspection of (\ref{browhzd}), the first order time derivative ensures
that $W$ will be a Brownian motion in time at fixed $x$, while the
introduction of the partial derivative with
respect to $x$ ensures the continuity of $W$ with respect to $x$.
The solution of (\ref{browhzd}) reads
\be
W(t,x) = \int_0^x dy \int_0^t dv ~\eta (v,y)~.
\label{solkjn}
\ee
This process (\ref{solkjn}) is known as the Brownian sheet\footnote{See
Walsh (1986).} and has the following covariance, which can be readily
obtained using (\ref{cvgherfv}) with
(\ref{solkjn})\,:
\be
\Cov \left[ W(t,x),W(t',x') \right] = (t \wedge t')~( x \wedge x')~,
\label{fondidkkkkk}
\ee
where $(t \wedge t')$ is the minimum of $t$ and $t'$.

For the calculus used above in the implementation of the
no-arbitrage condition on forward rate dynamics, we need the correlation of the
time increments of the string. The covariance is the
limit when $\delta t \to dt$ of
$\Cov \left[ W(t+\delta t, x) - W(t,x),W(t+\delta t, x') - W(t,x')
\right]$. Its calculation reduces to that of four terms of the form
(\ref{fondidkkkkk}) that simplify to
\be
\Cov \left[ d_t W(t,x),d_t W(t,x') \right] = (x \wedge x')~dt~.
\ee
or, stated in terms of the correlation,
\be
c(t,x,x') \equiv \Corr \left[ d_t W(t,x),d_t W(t,x') \right] = (x \wedge x')~.
\label{deffedde}
\ee

The property that the variance of $W(t,x)$ is not homogeneous in $x$, makes
this process unsatisfactory as a stochastic perturbation in
(\ref{modelsheet}): the instantaneous spot rate ($x=0$)
would have no volatility in this model since $W(t,0)=0$, and the variance of
forward rates would increase linearly
with $x$.

A simple modification of the Brownian sheet 
can be obtained by making $j(x) = x$ in (\ref{gindep}), leading to
\be
d_t \hat{W}(t,x) = dt~ {1 \over \sqrt{x}} \int_0^{x} dy ~\eta(t,y)~.
\ee
This process is just like the Brownian sheet, however rescaled by the factor
${1 \over \sqrt{x}}$ to ensure homogeneity in $x$. This form could have
been guessed
directly from the expression (\ref{fondidkkkkk}) of the correlation of the
Brownian sheet. The corresponding SPDE is
\be
{\partial \over \partial x} \biggl(\sqrt{x} {\partial \hat{W}(t,x) \over
\partial t}\biggl)
= \eta(t,x)~,
\ee
and
\be
\Corr \left[ d_t \hat{W}(t,x),d_t \hat{W}(t,x') \right] = 
\sqrt{x \over x'}~, \label{prqdhtyjjl}
\ee
for $x < x'$ and $\sqrt{x' \over x}$ for $x > x'$.

\subsection{The Ornstein-Uhlenbeck sheet}

A natural model for $Z(t,x)$ is one similar to
the Brownian sheet but in
which the variance is the same for all times-to-maturity. We look for the
string with the same variance for all $x$
and covariance between $x$ and $x'$ that depends only on the difference
$(x-x')$.

The O-U string satisfies (\ref{gindep}) with $j(x) = e^{2\kappa x}$, which
leads to
\be
U(t,x) = e^{-\kappa x} W(t,e^{2\kappa x}) = e^{-\kappa x}
\int_0^{e^{2\kappa x}} dy \int_0^t dv ~\eta(v,y)~.
\label{defzzzz}
\ee

The corresponding SPDE is
\be
{\partial \over \partial x} \biggl(\sqrt{j(x)} {\partial Z(t,x)
\over \partial t}\biggl)
= \sqrt{d j(x)\over dx} ~\eta(t,x)~.
\ee
which we can simplify to
\be
{\partial^2 U(t, x) \over \partial x \partial t} +
\kappa {\partial U(t, x) \over \partial t} = \sqrt{2\kappa} ~\eta(t,
x)~.
\label{rrrffff}
\ee

By inspection of
(\ref{defzzzz}), we get the following properties
\be
\Cov \left[ U(t,x),U (t',x') \right] = (t \wedge t') ~e^{-\kappa
|x-x'|}~,
\label{prporjdl}
\ee
and
\be
c(t,x,x')\equiv \Corr \left[ d_t U(t,x),d_{t}U(t,x') \right] = e^{-\kappa
|x-x'|}~.
\label{erfdvxcb}
\ee

Expressions (\ref{prporjdl}) and (\ref{erfdvxcb}) both show that, for fixed
$x$ ($x=x'$),
$U(t,x)$ is a Brownian motion in time with variance independent
of $x$. This is
the property that we were looking for, so that all times-to-maturity have
a priori
the same volatility for the driving stochastic process.\footnote{Different
volatilities for forward rates with different times-to-maturity can be
obtained
by choice of the weighting factor
$\sigma(t,x)$ in (\ref{modelsheet}).} We also learn that the time
increments
$d_t U(t,x)$ and $d_{t}U(t,x')$ are
correlated accross times-to-maturity\,: this is necessary to obtain
the continuity condition along $x$. Again, if the variations along $x$ were
delta-correlated, the process would consist of independent one-dimensional Brownian motions, one
for each $x$, thus leading to an almost everywhere discontinuous process. The
correlations along $x$ are the strongest between close times-to-maturity
and vanish
exponentially between distant times-to-maturity.
The parameter $\kappa$ allows us to span
a set of models, from
simple Brownian motion to a stochastic string. Indeed, for $\kappa \to 0$, all
times-to-maturity are so strongly coupled that they all consist of the same
single dimensional Brownian motion. This limit $\kappa \to 0$
thus recovers the single factor model driving simultaneously all
times-to-maturities. Figure 1 shows an example of the increment
in the O-U sheet for different values of $\kappa$. We see that, as $\kappa$
increases, the shape of the string has less and less humps.

There is another way to construct the O-U sheet.
Let us come back to the definition (\ref{defzzzz}) of $U(t,x)$ and
make the change of variable $y \to e^{2\kappa z}$ in the integral. We get
\be
U(t,x) = 2\kappa e^{-\kappa x}
\int_{-\infty}^{x} dz e^{2\kappa z} \int_0^t dv ~\eta(v,e^{2\kappa z})~.
\label{defzdfvsvzz}
\ee
We can write $\eta(v,e^{2\kappa z}) = {e^{-\kappa z} \over \sqrt{2\kappa} }
{\hat \eta}(t,z)$,
where ${\hat \eta}(t,z)$ is still white noise
$\Cov \left[ {\hat \eta}(t,z),{\hat \eta}(t',z') \right] =
\delta(t-t')~\delta(z-z')$. Dropping the hat
on the noise, we get
\be
U(t,x) = \sqrt{2\kappa} e^{-\kappa x}
\int_{-\infty}^{x} dz ~e^{\kappa z} \int_0^t dv ~\eta(v,z)~.
\label{defssccxv}
\ee
We then write $\int_{-\infty}^{x} = \int_{-\infty}^{0} + \int_{0}^{x}$. The
first integral gives a contribution
\be
e^{-\kappa x} \int_0^t dv \int_{-\infty}^{0} dz~ e^{\kappa z} \sqrt{2\kappa}
~\eta(v,z)
= e^{-\kappa x} \int_0^t dv  ~\eta(v)~,
\ee
where $\eta(v) \equiv \sqrt{2\kappa} \int_{-\infty}^0 dz~e^{\kappa z}
\eta(v,z)$ is
a simple Brownian process of unit variance. We have kept the same
notation $\eta$, but
it is clear that it now depends only on time. This is thus the noise
process that is
common to all times-to-maturity. If considered alone, this would recover
the single factor model driving simultaneously all times-to-maturity.
Putting everything together, we obtain the alternative formulation of the O-U
sheet
\be
U(t,x) = e^{-\kappa x} \int_0^t dv  ~\eta(v) +
\sqrt{2\kappa}~ e^{-\kappa x} \int_0^t dv
\int_{0}^{x} dz~ e^{\kappa z} ~\eta(v,z)~.
\label{defssccgcvcvxv}
\ee
This form exemplifies the relationship with the standard
one-dimensional O-U process, in the sense
that it shows the exponential dependence of the process on innovations
at different times-to-maturity. This is thus the generalization in the
time-to-maturity dimension of the classical O-U process usually defined
in the time direction.
The analogy with the one-dimensional O-U
process is also
clear from the SPDE formulation (\ref{rrrffff}) where it is seen that
the exponential
dependence is reflected in the mean reversion term ${\partial U \over
\partial t}$, which plays a role similar to the mean reversion term $-\kappa U$
in (\ref{ejjfppoi}).

This process, written in the form (\ref{defssccgcvcvxv}),
has been used by Kennedy (1997) and Goldstein (1997) to model forward
rates.\footnote{Kennedy (1997) only considers the case of a constan
volatility
function for the forward rates, whereas Goldstein (1997) allows for a general
volatility function. In contrast to our approach, both authors model forward
rates of fixed maturity date.} Our approach clarifies
the special role played by this process and its relationship with SPDE's.

\subsection{The integrated Ornstein-Uhlenbeck sheet}

As an interesting example, introduce the ``integrated O-U sheet'', defined by
\be
V_1(t,x) = \int_{0}^x dy ~U(t,y)~,
\label{ervcpo}
\ee
where $U(t,y)$ is the O-U sheet defined in (\ref{defzzzz}).
By definition, ${\partial V_1(t,x) \over \partial x} = U(t,x)$.
Since $U(t,x)$ is continuous in $x$, ${\partial V_1(t,x) \over
\partial x}$ is also continuous in $x$ and $V_1(t,x)$ is thus differentiable
with respect to $x$.

Since $U(t,y)$ satisfies the SPDE (\ref{rrrffff}), we see, by replacing
$U(t,y)$ by ${\partial V_1(t,x) \over \partial x}$ in this equation, that
$V_1(t,x)$ is the solution of the following SPDE\,
\be
{\partial^3 V_1(t, x) \over \partial x^2 \partial t} +
\kappa {\partial^2 V_1(t, x) \over \partial x \partial t}
= \sqrt{2\kappa} ~\eta(t, x)~.
\label{rrrffdcvsfff}
\ee
This is a SPDE of order three.

We obtain
$$
c(t,x,x')\equiv \Corr \left[ d_t V_1(t,x),d_t V_1(t,x') \right] = \int_{0}^x
dy
\int_{0}^{x'} dy' ~\Corr \left[ U(t,y),~U(t,y') \right]
$$
\be
= {1 \over \kappa} ~\biggl( 2(x \wedge x') - {1 \over \kappa}
(1-e^{-\kappa x})(1-e^{-\kappa x'}) \biggl)~,
\label{rdsccxxx}
\ee
where we have used (\ref{erfdvxcb}).

Inspired by the alternative formulation (\ref{defssccgcvcvxv})
of the O-U sheet $U(t,y)$
in terms of the exponential kernel, we define
another integrated O-U sheet $V_2(t,x)$ by
\be
V_2(t,x) = \kappa \sqrt{2} e^{-\kappa x} \int_{0}^x dy ~e^{\kappa y}~U(t,y)~,
\label{ervcpofvx}
\ee
Again by construction, ${\partial V_2(t,x) \over \partial x}$ is continuous
and
thus $V_2(t,x)$ is differentiable in $x$. By differentiation of
(\ref{ervcpofvx})
with respect to $x$, we get
\be
{\partial V_2(t,x) \over \partial x} = -\kappa V(t,x) + \kappa \sqrt{2}~U
(t,x)~.
\ee
The SPDE obeyed by $V_2(t,x)$ is simply
obtained by putting $U(t,x) =
{1 \over \kappa \sqrt{2}} [{\partial V_2(t,x) \over \partial x} + \kappa
V_2(t,x)]$ in
(\ref{rrrffff}). This is again a parabolic SPDE of order three.

The correlation of the time increments of this process looks nicer than
(\ref{rdsccxxx})\,:
$$
c(t,x,x')\equiv \Corr \left[ d_t V_2(t,x),d_t V_2(t,x') \right]
$$
\be
= 2 \kappa^2 ~e^{-\kappa (x+x')}
\int_{0}^x dy
\int_{0}^{x'} dy' ~e^{\kappa (y+y')}~\Corr \left[ d_tU(t,y),~d_t U (t,y')
\right]
=(1+ \kappa|x-x'|) e^{-\kappa |x-x'|}~.
\label{dfsccxdddw}
\ee
Figure 2 shows an example of the increment
in the integrated O-U sheet for different values of $\kappa$. We see that
the string is always smooth and, as $\kappa$
increases, its shape has less humps. 

This process
has been independently used by Goldstein (1997) to model forward rates.
Our approach shows that it is
one particular example among an arbitrarily large class of processes
exhibiting similar properties.

\subsection{The term structure of correlations}

Empirical observation shows that
correlation between
two forward rates, with maturities separated by a given interval, increases
with maturity. What are the string processes that exhibit this property?
Let us start from the parametrization obtained by assuming that
$h(t,x,y)$ is independent of $y$, which leads to (\ref{rfdscxw}).
Consider two time-to-maturities $x$ and $x' = x + \delta x$, where
$\delta x > 0$ is small compared to $x$. We can then expand
$j(x') = j(x) + \delta x {dj(x) \over dx} ~+ $ higher order terms.
Expanding the square root in (\ref{rfdscxw}), we get
\be
c(t,x,x') \equiv \Corr \left[ d_t Z(t,x), d_t Z(t,x') \right] \approx
 \biggl( 1 - {\delta x \over 2} {d\log j(x) \over dx}\biggl) ~.
\label{rfdsiiiicxw}
\ee
We see that $\Corr \left[ d_t Z(t,x), d_t Z(t,x') \right]$ increases with $x$
if ${d\log j(x) \over dx}$ decreases (while remaining positive) as $x$ increases.
An interesting situation is when ${d\log j(x) \over dx} \to 0^+$ as $x$
gets large.
In this case, the increments become perfectly correlated
for large time-to-maturities.

Parametrizing $j(x) = e^{i(x)}$, the above condition implies that $i(x)$ must be
a function which increase slower than $x$, in such a way that its derivative
decreases with $x$, that is, 
$i(x)$ must be concave.
In order words, $j(x)$ must grow slower than an exponential. This is not 
the case of the O-U sheet or the integrated O-U sheet.
However, the previous example (\ref{prqdhtyjjl}), where $j(x) = x$, qualifies. 
In fact, any expression like
$j(x) = e^{x^{\alpha}}$ with an exponent $0 < \alpha < 1$ qualifies. This
corresponds to  
\be c(t,x,y) \equiv \Corr \left[ d_t Z(t,x), d_t Z(t,y) \right] = 
e^{-\kappa |x^{\alpha}-y^{\alpha}|} ~. \label{corralpha} \ee

For instance, take $j(x) = e^{2\kappa \sqrt{x}}$. This corresponds to
a correlation function
\be c(t,x,y) \equiv \Corr \left[ d_t Z(t,x), d_t Z(t,y) \right] = 
e^{-\kappa|\sqrt{x}-\sqrt{y}|} \label{corrsqrt} \ee
which again works fine. Figure 3 shows the correlations 
between forward rates at maturities spaced by different time intervals, 
as the maturies increase. We see that correlation increases both with the
proximity of the maturities of the forward rates and with the 
times-to-maturity of the forward rates.

These models make it very easy to fit any covariance matrix of 
the increments of instantaneous forward rates. 
The function $\sigma$ can be used to fit the variances, and
the parameters $\kappa$ in (\ref{corrsqrt}) or $\kappa$ and $\alpha$ in
(\ref{corralpha}) can be used to fit the correlations.


\section{Empirical Implications}


Traditional factor models of the 
term structure, are only compatible with samples that include
at most as many bonds as there are factors in the model. In order to
be able to use data on more bond prices to estimate
these models, it is necessary to
add error terms to their econometric specification.
For econometric tractability, these errors must be assumed to be independent 
from the factors.\footnote{See Pearson and Sun (1994) or Chen and
Scott (1993).}  
This approach is correct if the incompatibility of the data with
the model is due to observation error, possibly due to bid-ask spreads
or nonsynchronous observations. However, if the incompatibility is
due to a misspecified model, the error terms will not be independent
of the factors and the econometric model will be inapropriate.

Models with stochastic string shocks can be estimated without
any such error terms. The models are compatible with any sample 
of forward rates (or bond prices) of finitely spaced maturities,
taken at some finite sampling interval. This is so, because, for
any parametric specification, there
is always a possible path for the string shock over a 
finite interval that can lead from the forward curve at
the begining of the interval to the forward curve at the end of
the interval. This realization may be highly unlikely, but
it is always possible. The estimation exercise is thus one 
of finding the most likely parameters, given a set of movements of
the forward curve over time.


\section{Option Pricing and Replication}


In our model, derivatives can in general be priced by simulation. 
We just need to simulate the dynamics of the forward curve under the risk adjusted
probability measure. Define a new string $Z^*$
with dynamics
\be
d_t Z^*(t,y)=d_t Z(t,y)-\int_0^{\infty}dy~ c(t,x,y) \phi(t,y)~.
\ee
Under the risk adjusted measure, this string is a martingale. Asset prices
discounted at the bank account are also martingales under this probability measure. 
Thus, the price at time $t$ of a (European, path-independent) contingent claim 
with payoff $\Phi(s)$ at time $s>t$ is
\be \Phi(t)=\E_t^* \left[ \Phi(s) e^{-\int_t^s du ~r(u)} \right] ~. \ee

It is sometimes convenient to use the $s$-maturity bond price as numeraire. 
In order to change drifts for the $s$-forward risk neutral
probability measure, note that
\be
v(t,s)=\int_0^{s-t}dx \int_0^{s-t}dy~ c(t,x,y) \sigma(t,x) \sigma(t,y)
\ee
is the instantaneous variance of the $s$-maturity bond price. Then, 
defining a new string $Z^s$ with dynamics
\be
d_t Z^s(t,y)=d_t Z^*(t,y) + v(t,s) ~ dt ~,
\ee
we can price the contingent claim as
\be \Phi(t)=P(t,s) ~ \E_t^s \left[ \Phi(s) \right] ~. \ee
Now the expectation is taken with respect to the probability measure under
which $Z^s$ is a martingale.

Unfortunately, we cannot use the Feynman-Kac Theorem to obtain a PDE for 
pricing contingent claims. In our model, there is no finite set of 
state variables that determine the value of interest rate derivatives.
Derivatives will in general depend on the full history of
(string) shocks to the forward rate curve. This shows that to hedge them,
it is necessary to use the full set of bonds of all maturities.

Closed form solutions for option prices can be obtained for Gaussian cases,
where the volatilities of forward rates are deterministic. Kennedy (1994) 
gives a formula for pricing European call options on bonds when the string 
used is the O-U sheet and $\sigma(t,x)=\sigma e^{- a x}$, with $\sigma$ and 
$a$ positive constants.


\section{Future Work}


In future work, we intend to estimate and test different parametrizations
of our model.
Another avenue of research that we are currently exploring is the
modelization of forward
rates directly as a stochastic string, i.e., as the solution of a SPDE,
rather
than the approach followed in the present paper of using stochastic strings
multiplied by volatility functions to shock forward rates. This new approach
is particularly promising for a parsimonious description of forward rate
curves with non-linear dynamics.

\hfill

\pagebreak


\section*{A \, Solution of Traditional Model}


In this appendix, we obtain the arbitrage-free process for forward rates of
the
traditional model of section \ref{simpmod}.

We perform a change of variable $(t,x) \to (\tau \equiv x+t, \xi
\equiv x-t)$ and denote ${\hat f}(\tau,\xi) \equiv f(t,x)$. We then have
\be
{\partial f(t,x)  \over \partial t} = {\partial\tau \over \partial t}
{\partial {\hat f}(\tau,\xi)  \over \partial \tau} +
{\partial\xi \over \partial t} {\partial {\hat f}(\tau,\xi)  \over \partial
\xi} =
{\partial {\hat f}(\tau,\xi)  \over \partial \tau} -
{\partial {\hat f}(\tau,\xi)  \over \partial \xi}~.
\ee
and, similarly,
\be
{\partial f(t,x)  \over \partial x} = {\partial\tau \over \partial x}
{\partial {\hat f}(\tau,\xi)  \over \partial \tau} +
{\partial\xi \over \partial x} {\partial {\hat f}(\tau,\xi)  \over \partial
\xi} =
{\partial {\hat f}(\tau,\xi)  \over \partial \tau} +
{\partial {\hat f}(\tau,\xi)  \over \partial \xi}~.
\ee
Replacing these in the l.h.s. of (\ref{zetr}), we get
\be
{\partial f(t,x)  \over \partial t} - {\partial f(t,x)  \over \partial x}
= - 2 {\partial {\hat f}(\tau,\xi)  \over \partial \xi}~.
\ee
Equation (\ref{zetr}) thus becomes
\be
{\partial {\hat f}(\tau,\xi)  \over \partial \xi} = -{1 \over 2}
a\left({\tau-\xi \over 2},{\tau+\xi \over 2}\right)  - {1 \over 2}
\sigma\left({\tau-\xi \over 2},{\tau+\xi \over 2}\right)
~ \left. {dW(t) \over dt} \right|_{t={\tau-\xi \over 2}}~,
\ee
where we have expressed the values of $t$ and $x$ in terms of the new
variables $\tau$ and $\xi$\,: $t = {\tau-\xi \over 2}$ and $x =
{\tau + \xi \over 2}$. This change of variable allows us to get an ODE from
the initial PDE and its straightforward integration gives
\be
{\hat f}(\tau,\xi) - {\hat f}(\tau,\xi_0) =  -{1 \over 2}
\int_{\xi_0}^{\xi} dw~ a\left({\tau-w \over 2},{\tau+w \over 2}\right)
- {1 \over 2} \int_{\xi_0}^{\xi} dW(w)
~ \sigma\left({\tau-w \over 2},{\tau+w \over 2}\right)  ~~.
\label{erertgk}
\ee

Now, by definition of the change of variables,
${\hat f}(\tau,\xi) = f(t,x)$. We have a free choice for $\xi_0$. Since we
want
to
express $f(t,x)$ as a function of $f(0,.)$, $\xi_0$ is chosen such that
it corresponds to $t_0=0$. To see what this leads to, we use the definition
${\hat f}(\tau,\xi_0) = f(t_0={\tau-\xi_0 \over 2},{\tau+\xi_0 \over 2})$.
For
$t_0$ to be zero, this implies that $\xi_0 = \tau$ and thus
${\tau+\xi_0 \over 2} = \tau = x+t$.
The l.h.s. of (\ref{erertgk}) thus reads
\be
{\hat f}(\tau,\xi) - {\hat f}(\tau,\xi_0)  = f(t,x) - f(0,x+t)
\label{bcv}
\ee
To tackle the r.h.s. of (\ref{erertgk}), we perform the change of variable
from
$w$ to $v = {\tau - w \over 2}$. Then ${\tau + w \over 2} = \tau - v$
and $\int_{\xi_0=\tau=x+t}^{\xi=x-t} dw = -2 \int_0^t dv$.
Finally, we get 
\be
f(t,x) = f(0,x+t)  + \int_0^t dv~ a(v,t+x-v)
+ \int_0^t dW(v) ~ \sigma(v,t+x-v)  ~.
\ee

Let us now comment on the validity of this derivation when $f(t,x)$ is not
differentiable in $x$
(and in $t$ as it is usually). As a teaching example, consider the SPDE
\be
{\partial W \over \partial t} - {\partial W \over \partial x} = \eta(t,x)~.
\label{zzzzaaa}
\ee
Performing the same change of variable as above, we get
\be
{\partial {\hat W}(\tau, \xi) \over \partial \xi} = -{1 \over 2} {\hat
\eta}(\tau,\xi)~,
\label{eexxxxg}
\ee
where
\be
{\hat \eta}(\tau,\xi) \equiv \eta((\tau-\xi)/2,(\tau+\xi)/2)~.
\ee
The solution of (\ref{eexxxxg}) reads
\be
{\hat W}(\tau,\xi) = -{1 \over 2} \int_0^{\xi} dy~{\hat \eta}(\tau,y)
= -{1 \over 2} \int_0^{x-t} dy~\eta((x+t-y)/2,(x+t+y)/2)~.
\label{soodhhgggg}
\ee

One readily checks that (\ref{soodhhgggg}) obeys (\ref{zzzzaaa}). The terms
formally involving the derivatives of the noise $\eta$ would take sense when
discretizing the equations.


\section*{B \, Requirements for String Shocks}


\subsection*{B.1 Martingale condition for strings}

We want $E_t[d_t X(t,x)]$ to be equal to zero.
From (\ref{dggvvgggg}), we have
\be
d_t X(t,x) = \int_0^t dv \int_{-\infty}^{\infty} dy
~{\partial G(t,x|v,y) \over \partial t}~ \eta(v,y) +
\int_{-\infty}^{\infty} dy ~G(t,x|t,y)~ \eta(t,y)~.
\label{dfgvxvvvv}
\ee
The expectation $E_t[d_t X(t,x)]$ conditional on the realization of
$X(t,x)=g(x)$, where $g(x)$ is a specified function, can be
written as
\be
E_t[d_t X(t,x)]  = \int...\int {\cal D}\{\eta\} {\cal P}(\{\eta\})
~d_t X(t,x)~ \delta \biggl( X(t,x) - g(x) \biggl)~,
\label{dhhdjdjdjjd}
\ee
where $\int...\int {\cal D}\{\eta\}$ denotes the functional
integral\footnote{To get an intuitive representation, imagine that time is
discrete. Then, the integrals are carried out over the variables $\eta_i$ for
time $i=1$ to $t$.}
over all possible realizations of the noise $\eta$, ${\cal P}(\{\eta\})$ is
the probability density function weighting
each realization of the noise, and the term $\delta \biggl( X(t,x) - g(x)
\biggl)$ ensures that the functional integral is
carried over all possible $\eta$'s {\it restricted} to satisfy the specific
realization of $X(t,x)$. With delta correlated
shocks $\eta$, ${\cal P}(\{\eta\})$ is Gaussian, so that
\be
{\cal P}(\{\eta\}) = P_0 ~e^{-\int dv \int dy~C^{-1}(v,y)~[\eta(v,y)]^2}~,
\label{pppoooii}
\ee
where $2C^{-1}(v,y)$ is the inverse of the variance of $\eta(v,y)$ and we
allow
it to vary with time and time-to-maturity.

Taking the Fourier transform of  $E_t[d_t X(t,x)]$  with respect to
$g(x)$ gives
\be
{\hat E}_t[d_t X(t,x,k)] = \int...\int {\cal D}\{\eta\} {\cal
P}(\{\eta\})
~d_t X(t,x)~ e^{ik~X(t,x)}~,
\label{dhhdjdjdjgdjd}
\ee
where $k$ is the conjugate of $g(x)$ in the Fourier transform. We notice
that the second term in the r.h.s. of (\ref{dfgvxvvvv}) does not contribute
to ${\hat E}_t[d_t X(t,x,k)]$ since it contains the innovations
$\eta(t,y)$ that are {\it posterior} to those contributing to $X(t,x)$. The
gaussian integrals over the centered innovations $\eta(t,y)$ thus vanish.

The integrand in (\ref{dhhdjdjdjgdjd}) contains the exponential term
$$
\exp \biggl(-\int dv \int dy~\biggl[C^{-1}(v,y)~[\eta(v,y)]^2 -
ik~G(t,x|v,y)~\eta(v,y)\biggl] \biggl)~.
$$
$C^{-1}(v,y)~[\eta(v,y)]^2 - ik~G(t,x|v,y)~\eta(v,y)$ can be factorized as
$$
C^{-1}(v,y)~[\eta(v,y)]^2 - ik~G(t,x|v,y)~\eta(v,y) =
$$
\be
C^{-1}(v,y)~\biggl(\eta(v,y) - ik~{G(t,x|v,y) \over 2C^{-1}(v,y)}\biggl)^2~
+ {k^2 ~[G(t,x|v,y)]^2 \over 4C^{-1}(v,y)}~.
\ee
The Gaussian integrals can be carried over ${\hat \eta}(v,y) =
\eta(v,y) - ik~{G(t,x|v,y) \over 2C^{-1}(v,y)}$, which leads to
\be
{\hat E}_t[d_t X(t,x,k)] = \int_0^t dv \int_{-\infty}^{+\infty} dy
~~{ik \over C^{-1}(v,y)}~{\partial [G(t,x|v,y)]^2 \over \partial t}~
e^{-{k^2 ~\int dv \int dy~[G(t,x|v,y)]^2 \over 4C^{-1}(v,y)}}~.
\label{terrccccxxx}
\ee
We look for the condition that the Green function $G(t,x|v,y)$ must satisfy
for this expression (\ref{terrccccxxx}) to be identically zero for any $t$
and $x$.\footnote{If the Fourier transform
is zero, the function is zero.}

The condition is simply that
\be
{\partial G(t,x|v,y) \over \partial t} = 0~,
\label{chjcchhh}
\ee
for all $t$ and $x$, i.e. $G(t,x|v,y)$ must be {\it independent} of $t$. We
note that the dependence in $t$ can still appear through dependence in time
of possible upper and lower bounds in
the integral over y.
We verify that this condition is observed for the parametric examples of
stochastic string shocks given in the text.

The condition (\ref{chjcchhh}) implies that the Fourier transform
(\ref{dhhdjdjdjgdjd})
of $E_t[d_t X(t,x)]$ is zero, and thus $E_t[d_t X(t,x)]$ is zero itself. This
is a Martingale condition.

\subsection*{B.2 Markov condition for strings}

An additional requirement (more stringent than the previous one) is that
$X(t,x)$ be Markovian,
i.e. that its increment $d_t X(t,x)$ be completely uncorrelated from past
time innovations.
The general form that the stochastic string can take in order to fulfill
this condition is
\be
X(t,x) = X(0,x) + \int_0^t dv~g(v) ~\int_{-\infty}^{j(x)} dy
~h(v,x,y) \eta(v,y)~,
\label{dzzzzcsgggg}
\ee
where $g(v)$, $j(x)$ and $h(v,x,y)$ are arbitrary functions. This expression
obeys the condition (\ref{chjcchhh}) and in addition time does not appear
in the bounds of the integral over $y$. As a consequence, from
(\ref{dzzzzcsgggg}),  we get
\be
d_t X(t,x) = dt~
g(t) ~\int_{-\infty}^{j(x)} dy ~h(t,x,y) \eta(t,y)~.
\label{dzzzzssd2}
\ee
The fact that $\eta(t,y)$ is immediately posterior to $t$ ensures that the
expectation of $d_t X(t,x)$ conditional to the realization of $X(t,x)$ is
identically zero.
This recovers a result of Kennedy (1997), obtained in a different setup.

\subsection*{B.3 Constraints on the covariance of the increments}

The variance $\Var \left[ d_t X(t,x) \right]$ of the
increments must not depend on $x$.
This condition requires some additional condition on the functions
$j(x)$ and $h(v,x,y)$ that we now derive. Notice that
$\Cov \left[ d_t X(t,x) d_t X(t,x') \right]$ does not
depend on $t$ if $g(t)$ is
constant.

From (\ref{dzzzzssd2}), we get
\be
\Var \left[ d_t X(t,x) \right] = dt~ ~\int_{0}^{j(x)} dy
~[h(t,x,y)]^2 ~,
\label{dzzzfdfzssd2}
\ee
putting $g(t) = 1$. We have modified the lower bound of the integral in $y$
so as not
to have to worry about additional constraints in order to ensure
convergence at $-\infty$.
The condition to impose is that (\ref{dzzzfdfzssd2}) be independent of $x$,
which leads to the solutions given in section 4.3.

\section*{C \, Differentiability in Time-to-Maturity}

The string processes studied in section 4.3 are continuous both in $t$ and
$x$ but
not differentiable either in $t$ or $x$. Of special interest is the
non-differentiability in $x$. Consider first the Brownian
sheet process (\ref{solkjn}) and let us consider ${\partial W(t,x) \over
\partial x} = \int_0^t \eta(v,x)$. This integral is a continuous
sum of random variables. The sum of Gaussian random variables is itself
Gaussian
 with variance equal to $t$. Thus, ${\partial
W(t,x) \over \partial x}$ is not a function but should rather be interpreted
as a noise process, discontinuous almost everywhere with respect to $x$.
The derivative with respect to $x$ does not exist strictly speaking, but
only in the sense that its integral is well defined. This
non-differentiability holds
true for a large class of noise processes, not necessarily Gaussian.

A similar reasoning holds for all the string processes defined in
(\ref{dzsdsd}).

All the stochastic partial differential equations that we discuss in this paper are linear.
For such linear equations, a particularly convenient mathematical tool is the
Fourier or Laplace transform. Let $Z(t,x)$ be one of the stochastic strings.
Its Fourier transform with respect to $x$ is
\be
{\hat Z} (t,k) \equiv \int_{-\infty}^{+\infty} dx ~e^{ikx} ~Z(t,x)~.
\label{Fiyuruer}
\ee
The inverse Fourier transform retrieving $Z(t,x)$ from ${\hat Z} (t,k)$ is
\be
Z(t,x) = \int_{-\infty}^{+\infty} dk ~e^{-ikx} ~{\hat Z} (t,k)~.
\label{Fiyuruererer}
\ee
We have the following useful property\,:
\be
\int_{-\infty}^{+\infty} dx ~e^{ikx} ~{\partial^p Z(t,x) \over \partial x^p} =
(-ik)^p~{\hat Z} (t,k)~,
\label{Fiyudggfdt}
\ee
for an arbitrary positive integer $p$ \footnote{Derivatives are taken in the
sense of distributions and we use the usual definition of the Fourier
transform
of distributions.}.
If $\eta(t,x)$ is a Gaussian white noise, we have
\be
\Cov \left[ {\hat \eta}(t,k) , {\hat \eta}(t',k') \right] = \delta (t-t')
~\delta(k-k')~.
\ee

Consider first the Brownian sheet (\ref{browhzd}). Taking the Fourier
transform of
(\ref{browhzd}) leads to
\be
{\partial {\hat W}(t,k) \over \partial t} = {i \over k} {\hat \eta} (t,k)~,
\label{rdfvccxwww}
\ee
whose solution is simply
\be
{\hat W}(t,k) = {\hat W}(0,k)  + {i \over k} \int_0^t d\tau ~{\hat \eta}
(\tau,k)~.
\ee
We thus get
\be
\Cov \left[ {\hat W}(t,k) {\hat W}(t',k') \right] = - {\delta(k-k') \over
k^2} ~({t \wedge t'})~.
\ee
This expression allows us to calculate any desired quantity. For instance,
we find
\be
\Cov \left[ {\partial W(t,x) \over \partial x} ,{\partial W(t',x') \over
\partial x'} \right]
= (t \wedge t')~\delta(x-x')~,
\ee
which shows that $\E \left[ \left|{\partial W(t,x) \over \partial
x}\right|^2 \right]$
is infinite,
the signature that $W(t,x)$ is not differentiable in $x$.

Consider now the O-U process (\ref{defzzzz}). Taking the Fourier transform
of the SPDE defining the O-U process, leads to
\be
(-ik + \kappa) {\partial {\hat Z} (t,k) \over \partial t} = \sqrt{2
\kappa}~{\hat \eta}(t,k)~ .
\ee
Its solution is
\be
{\hat Z} (t,k)  = {\hat Z} (0,k) + {\sqrt{2 \kappa} \over -ik + \kappa}
\int_0^t d\tau ~{\hat \eta}(\eta,k)~.
\ee
Following the same steps as before, we get
\be
\Cov \left[ {\partial Z(t,x), \over \partial x} {\partial Z(t',x') \over
\partial x'} \right]
= (t \wedge t')~\int_{-\infty}^{+\infty} dk ~e^{ik(x-x')} ~{2\kappa~k^2
\over k^2 + \kappa^2}~.
\ee
The integral still diverges for $|k| \to \infty$, again expressing the
non-differentiability of
$Z(t,x)$.

In contrast, the integrated process (\ref{ervcpo}) which obeys the SPDE
(\ref{rrrffdcvsfff}) has
its Fourier transform ${\hat Y}(t,k)$ satisfies
\be
(k^2 + ik \kappa) {\partial {\hat Y}(t,k) \over \partial t} = -
\sqrt{2\kappa}~{\hat \eta}(t,k)~.
\ee
The solution is
\be
{\hat Y}(t,k) = {\hat Y}(0,k) - {\sqrt{2\kappa} \over k^2 + ik \kappa}
\int_0^t d\tau~{\hat \eta}
(\tau, k)~,
\ee
leading to
\be
\Cov \left[ {\hat Y}(t,k) ,{\hat Y}(t',k') \right] = ({t \wedge t'})~ {2
\kappa \delta(k-k')
 \over k^4 + k^2~\kappa^2}~.
\ee
We thus see that
\be
\Cov \left[ {\partial Y(t,x) \over \partial x}~,~ {\partial Y(t',x') \over
\partial x'} \right]
= 2 \kappa~ ({t \wedge t'}) \int_{-\infty}^{+\infty} dk~{e^{ik(x-x')} \over
k^2 + \kappa^2} =
\pi~ ({t \wedge t'}) ~e^{-\kappa |x-x'|}~,
\ee
which is finite, signaling as expected the differentiability of $Y(t,x)$
with respect to $x$.

\pagebreak

\section*{References}

Bender, C.M. and S.A. Orszag, 1978, Advanced mathematical methods for
scientists and engineers, International Student Edition, McGraw-Hill.

Brace, A. and M. Musiela, 1994, ``A Multifactor Gauss Markov
Implementation of Heath, Jarrow, Morton'', {\it Mathematical Finance} 4,
259-283.

Brace, A., D. Gatarek and M. Musiela, 1997, ``The Market Model
of Interest Rate Dynamics'', {\it Mathematical Finance}, 7, 127-155.

Chen, R.-R., and L. Scott, 193, ``ML estimation for a multifactor
equilibrium model of the term structure'', {\it Journal of Fixed Income},
December.

Cox, J.C., J.E. Ingersoll and S.A. Ross, 1985, ``A Theory of the Term
Structure of Interest Rates'', {\it Econometrica} 53, 385-407.

Duffie, D., 1996, {\it Dynamic Asset Pricing Theory}, Second Edition, Princeton University Press.

Dybvig, P., J. Ingersoll and S. Ross, 1996, ``Long Forward and Zero-Coupon
Rates Can Never Fall'', {\it Journal of
Business} 69, 1-26.

Goldstein, R., 1997, ``The Term Structure of Interest Rates as a Random
Field'',
Preprint, Ohio State University.

Heath, D., R.A. Jarrow and A. Morton, 1992, ``Bond Pricing
and the Term Structure of Interest Rates: A New Methodology for Contingent
Claims Valuation'', {\it
Econometrica} 60, 77-105.

Jeffrey, A., 1997, ``Asymptotic Maturity Behavior of Single Factor
Heath-Jarrow-Morton Term Structure Models:
A Note'', Preprint, Yale School of Management.

Kennedy, D.P., 1994, ``The Term Structure of Interest Rates as a Gaussian Random Field'', 
{\it Mathematical Finance} 4, 247-258.

Kennedy, D.P., 1997, ``Characterizing Gaussian Models of the Term Structure
of Interest Rates'', {\it Mathematical Finance} 7, 107-118.

Miltersen, K., K. Sandmann and D. Sondermann, 1997, ``Closed
Form Solutions for Term Structure Derivatives with Log-Normal Interest
Rates'', {\it Journal of Finance}, 52, 409-430.

Morse, P.M. and H. Feshbach, 1953, {\it Methods in Theoretical Physics},
McGraw-Hill.

Musiela, M., 1993, ``Stochastic PDEs and Term Structure Models'',
Preprint.

Pearson, N.D. and T.S. Sun, 1994, ``Exploiting the Conditional Density in
Estimating the Term Structure: An Application to the Cox, Ingersoll, and
Ross Model'', {\it Journal of Finance} 49, 1279-304.

Vasicek, O., 1977, ``An Equilibrium Characterization of the Term Structure'', 
{\it Journal of Finance} 5, 177-188.

Walsh, J.B., 1986, in {\it Ecole d'Ete de Probabilite de Saint-Flour},
``An Introduction to Stochastic Partial Differential Equations'',
Springer-Verlag.

\pagebreak

FIGURE CAPTIONS

\vskip 1cm
Figure 1 \,: Sample increment 
to the O-U sheet, with correlation function (\ref{erfdvxcb}) 
for different values of the parameter $\kappa$. The time increment 
is taken to be 1, measured in the same units of time-to-maturity.

\vskip 1cm
Figure 2 \,: Sample increments to the integrated 
O-U sheet, with correlation function (\ref{dfsccxdddw}) for different
values of the parameter $\kappa$. The time increment is taken to be 1, 
measured in the same units of time-to-maturity.

\vskip 1cm
Figure 3 \,: Correlations between forward rates 
with times-to-maturity separated by different intervals, for the
string shock with correlation function (\ref{corrsqrt}). The parameter
$\kappa$ is taken to be 1.

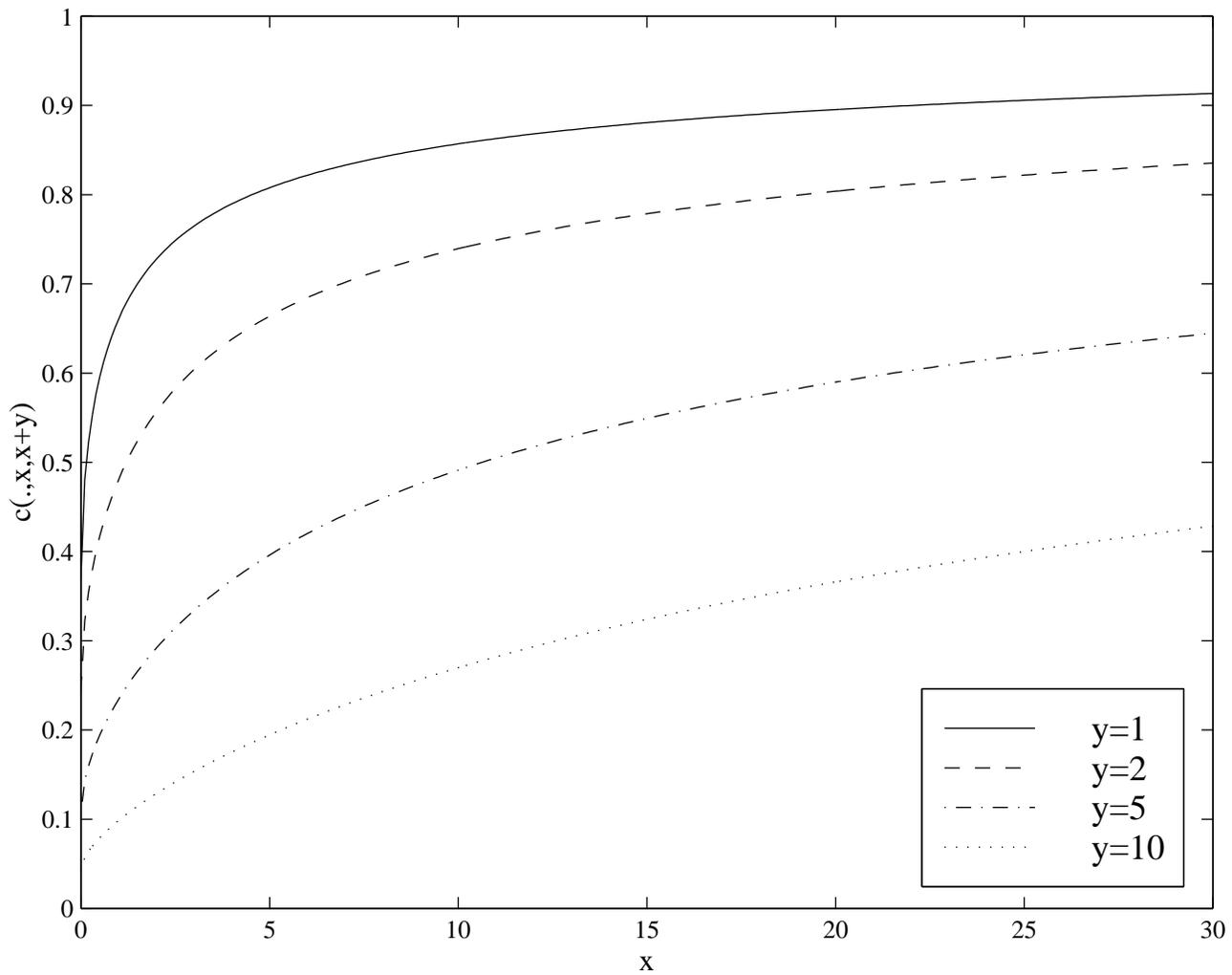
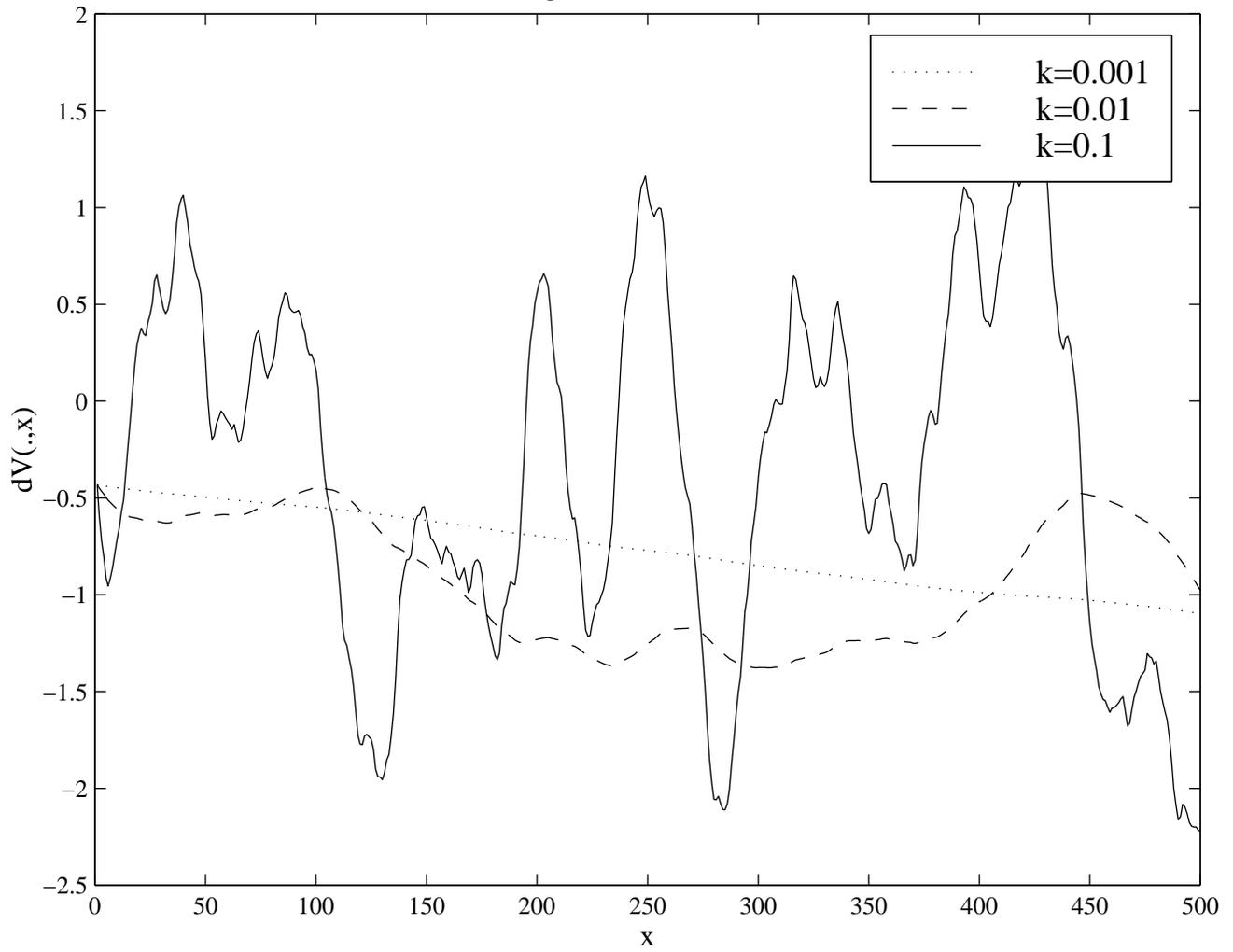
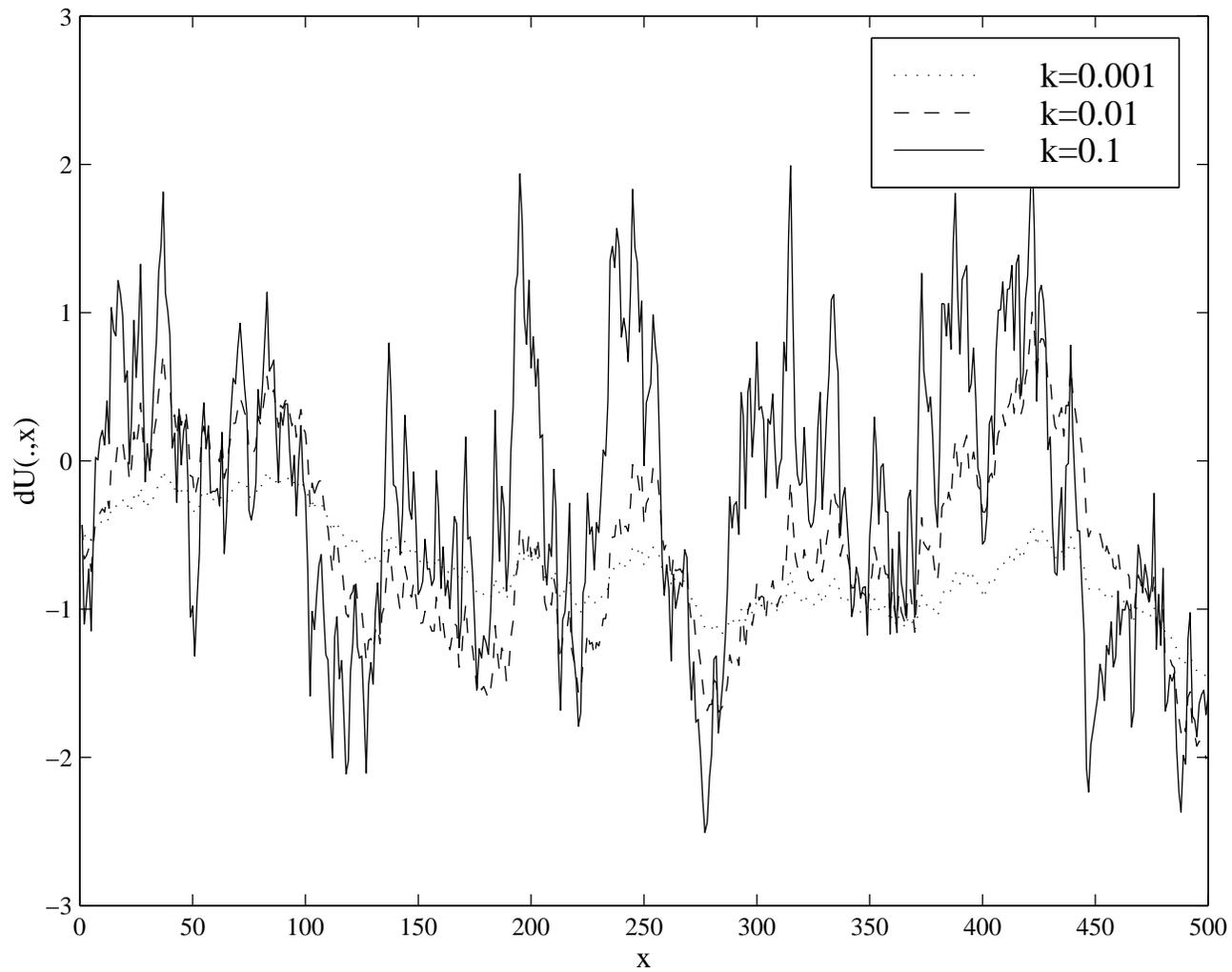
\end{document}